\documentstyle[12pt]{article}
\begin{document}

\begin{center}
{\large {\bf Superluminal motions?\\[0pt]
A bird-eye view of the experimental situation$^{\: (\dag)}$}} 
\footnotetext{$^{\: (\dag)}$ E-mail for contacts:  recami@mi.infn.it }
\end{center}

\vspace*{5mm}

\centerline{ Erasmo Recami }

\vspace*{0.5 cm}

\centerline{{\em Facolt\`{a} di Ingegneria, Universit\`{a} Statale di Bergamo,
Dalmine (BG), Italy;}} \centerline{{\em INFN---Sezione di Milano, Milan,
Italy; \ and}} \centerline{{\em CCS, State University of Campinas, Campinas,
S.P., Brazil.}}

\vspace*{1. cm}

\ 

\ 

{\bf 1. - Introduction.}\hfill\break

\hspace*{5ex} The question of Super-luminal ($V^{2}>c^{2}$) objects or waves
has a long story, starting perhaps in 50 b.C. with Lucretius' {\em De Rerum
Natura} (cf., e.g., book 4, line 201: [$<<$Quone vides {\em citius} debere
et longius ire/ Multiplexque loci spatium transcurrere eodem/ Tempore {\em %
quo Solis} pervolgant {\em lumina} coelum?$>>$]). \ Still in
pre-relativistic times, one meets various related works, from those by
J.J.Thomson to the papers by the great A.Sommerfeld. \ With Special
Relativity, however, since 1905 the conviction spread over that the speed $c$
of light in vacuum was the {\em upper} limit of any possible speed. For
instance, R.C.Tolman in 1917 believed to have shown by his ``paradox'' that
the existence of particles endowed with speeds larger than $c$ would have
allowed sending information into the past. Such a conviction blocked for
more than half a century ---aside from an isolated paper (1922) by the
Italian mathematician G.Somigliana--- any research about Superluminal
speeds. Our problem started to be tackled again essentially in the fifties
and sixties, in particular after the papers[1] by E.C.George Sudarshan et
al., and later on[2] by E.Recami, R.Mignani, et al. [who rendered the
expressions subluminal and Superluminal of popular use by their works at the
beginning of the Seventies], as well as by H.C.Corben and others (to confine
ourselves to the {\em theoretical} researches). \ The first experiments
looking for tachyons were performed by T.Alv\"{a}ger et al. 

\hspace*{5 ex} Superluminal objects were called tachyons, T, by G.Feinberg,
from the Greek word $\tau \alpha \chi {\acute{\upsilon}} \varsigma$, quick,
and this induced us in 1970 to coin the term bradyon, B, for ordinary
subluminal ($v^2<c^2$) objects, from the Greek word $\beta \rho \alpha
\delta {\acute{\upsilon}} \varsigma$, slow). At last, objects travelling
exactly at the speed of light are called ``luxons".

\hspace*{5ex} In recent years, terms as ``tachyon'' and ``superluminal''
fell unhappily into the (cunning, rather than crazy) hands of
pranotherapists and mere cheats, who started squeezing money out of
simple-minded people; for instance by selling plasters (!) that should cure
various illnesses by ``emitting tachyons''... \ We are dealing with them
here, however, since at least four different experimental sectors of physics
seem to indicate the actual existence of Superluminal motions [it is an old
use of ours to write Superluminal with a capital S], thus confirming some
long-standing theoretical predictions[3]. \ So much so that even the
N.Y.Times commented on May 30, 2000, upon two of such experiments, imitated
the next day (and again at the end of the next July) by nearly all the world
press. \ In this rapid informative paper, after a sketchy theoretical
introduction, we are setting forth a reasoned outline of the experimental
state-of-art: brief, but accompanied by a bibliography sufficient in some
cases to provide the interested readers with coherent, adequate information;
and without forgetting to call attention ---at least in the two sectors more
after fashion today--- to some other worthy experiments.

\ 

{\bf 2. Special and Extended Relativity}.\hfill\break

\hspace*{5 ex} Let us premise that special relativity (SR), abundantly
verified by experience, can be built on two simple, natural Postulates: \ 1)
that the laws (of electromagnetism and mechanics) be valid not only for a
particular observer, but for the whole class of the ``inertial" observers: \
2) that space and time be homogeneous and space be moreover isotropic. \
From these Postulates one can theoretically {\em infer} that one, and only
one, {\em invariant} speed: and experience tells us such a speed to be that, 
$c$, of light in vacuum; in fact, light possesses the peculiar feature of
presenting always the same speed in vacuum, even when we run towards or away
from it. \ It is just that feature, of being invariant, that makes quite
exceptional the speed $c$: no bradyons, and no tachyons, can enjoy the same
property!

\hspace*{5ex} Another (known) consequence of our Postulates is that the
total energy of an ordinary particle increases when its speed $v$ increases,
tending to infinity when $v$ tends to $c$. Therefore, infinite forces would
be needed for a bradyon to reach the speed $c$. This fact generated the
popular opinion that speed $c$ can be neither achieved nor overcome. \
However, as speed $c$ photons exist which are born live and die always at
the speed of light (without any need of accelerating from rest to the light
speed), so particles can exist ---tachyons[4]--- always endowed with speeds $%
V$ larger than $c$ (see Fig.1). \ This circumstance has been picturesquely
illustrated by George Sudarshan (1972) with reference to an imaginary
demographer studying the population patterns of the Indian subcontinent: $<<$%
Suppose a demographer calmly asserts that there are no people North of the
Himalayas, since none could climb over the mountain ranges! That would be an
absurd conclusion. People of central Asia are born there and live there:
they did not have to be born in India and cross the mountain range. So with
faster-than-light particles$>>$. \ Let us add that, still starting from the
above two Postulates (besides a third one, even more obvious), the theory of
relativity can be generalized[3,4] in such a way to accommodate also
Superluminal objects; such an extension is largely due to the Italian
school, by a series of works performed mainly in the Sixties--Seventies. \
Also within the ``Extended Relativity''[3] the speed $c$, besides being
invariant, is a limiting velocity: but every limiting value has two sides,
and one can a priori approach it both from the left and from the right.

\hspace*{5ex} Actually, the ordinary formulation of SR is restricted too
much. For instance, {\em even leaving tachyons aside}, it can be easily so
widened as to include {\em antimatter\/}[5]. Then, one finds space-time to
be a priori populated by normal particles P (which travel forward in time
carrying positive energy), {\em and} by dual particles Q ``which travel
backwards in time carrying negative energy''. The latter shall appear to us
as {\em antiparticles}, i.e., as particles ---regularly travelling forward
in time with positive energy, but--- with all their ``additive'' charges
(e.g., the electric charge) reversed in sign!: see Fig.2. \ To clarify this
point, let us recall that we, macroscopic observers, have to move in time
along a single, well-defined direction, to such an extent that we {\em cannot%
} even see a motion backwards in time...; and every object like Q,
travelling backwards in time (with negative energy), will be {\em necessarily%
} reinterpreted by us as an anti-object, with opposite charges but
travelling forward in time (with positive energy).[3-5]

\hspace*{5 ex} But let us forget about antimatter and go back to tachyons. A
strong objection against their existence is based on the opinion that by
tachyons it be possible to send signals into the past, owing to the fact
that a tachyon T which ---say--- appears to a first observer $O$ as emitted
by A and absorbed by B, can appear to a second observer $O^{\prime}$ as a
tachyon T' which travels backwards in time with negative energy. However, by
applying (as it is obligatory to do) the same ``reinterpretation rule" or
switching procedure seen above, T' will appear to the new observer $%
O^{\prime}$ just as an antitachyon ${\overline{{\rm T}}}$ emitted by B and
absorbed by A, and therefore travelling forward in time, even if in the
contrary {\em space} direction. In such a way, every travel towards the
past, and every negative energy, do disappear...

Starting from this observation, it is possible to solve[5] the so-called
causal paradoxes associated with Superluminal motions: paradoxes which
result to be the more instructive and amusing, the more sophisticated they
are; \ but that cannot be re-examined here (some of them having been
proposed by R.C.Tolman, J.Bell, F.A.E.Pirani, J.D.Edmonds and others).[6,3]
\ Let us only mention here the following. \ The reinterpretation principle
---according to which, in simple words, signals are carried only by objects
which appear to be endowed with positive energy--- does eliminate any
information transfer backwards in time, but this has a price: That of
abandoning the ingrained conviction that the judgement about what is cause
and what is effect be independent of the observer. In fact, in the case
examined above, the first observer $O$ considers the event at A t be the
cause of the event at B. \ By contrast, the second observer $O^{\prime}$
will consider the event at B as causing the event at A. \ All the observers
will however see the cause to happen {\em before} its effect.

\hspace*{5ex} Taking new objects or entities into consideration always
forces us to a criticism of our prejudices. If we require the phenomena to
obey the {\em law} of (retarded) causality with respect to all the
observers, then we cannot demand also the phenomena {\em description}
``details'' to be invariant: namely, we cannot demand in that case also the
invariance of the ``cause'' and ``effect'' labels.[6,2] \ To illustrate the
nature of our difficulties in accepting that e.g. the parts of cause and
effect depend on the observer, let us cite an analogous situation that does
not imply present-day prejudices: $<<$For ancient Egyptians, who knew only
the Nile and its tributaries, which all flow South to North, the meaning of
the word ``south'' coincided with the one of ``upstream'', and the meaning
of the word ``north'' coincided with the one of ``downstream''. When
Egyptians discovered the Euphrates, which unfortunately happens to flow
North to South, they passed through such a crisis that it is mentioned in
the stele of Tuthmosis I, which tells us about {\em that inverted water that
goes downstream (i.e. towards the North) in going upstream}$>>$ (Csonka,
1970).

\hspace*{5ex} The last century theoretical physics led us in a natural way
to suppose the existence of various types of objects: magnetic monopoles,
quarks, strings, tachyons, besides black-holes: and various sectors of
physics could not go on without them, even if the existence of none of them
is certain (also because attention has not yet been paid to some links
existing among them: e.g., a Superluminal electric charge is expected to
behave as a magnetic monopole; and a black-hole a priori can be the source
of tachyonic matter). \ According to Democritus of Abdera, everything that
was thinkable without meeting contradictions had to exist somewhere in the
unlimited universe. This point of view ---which was given by M.Gell-Mann the
name of ``totalitarian principle''--- was later on expressed (T.H.White) in
the humorous form ``Anything not forbidden is compulsory''. Applying it to
tachyons, Sudarshan was led to claim that if tachyons exist, they must to be
found; if they do not exist, we must be able to say clearly why...\newline

\ 

{\bf 3. The experimental state-of-the-art}.\hfill\break

\hspace*{5 ex} Extended Relativity can allow a better understanding of many
aspects also of {\em ordinary} relativistic physics, even if tachyons would
not exist in our cosmos as asymptotically free objects. \ As already said,
we are dealing with them ---however--- since their topic is presently
returning after fashion, especially because of the fact that at least three
or four different experimental sectors of physics seem to suggest the
possible existence of faster-than-light motions. \ We wish to put forth in
the following some information (mainly bibliographical) about the
experimental results obtained in each one of those different physics sectors.%
\newline

\ 

{\bf A)} \ {\bf Neutrinos} -- First: A long series of experiments, started
in 1971, seems to show that the square ${m_{0}}^{2}$ of the mass $m_{0}$ of
muonic neutrinos, and more recently of electronic neutrinos too, is
negative; which, if confirmed, would mean that (when using a na\"{i}ve
language, commonly adopted) such neutrinos possess an ``imaginary mass'' and
are therefore tachyonic, or mainly tachyonic.[7,3] \ [In Extended
Relativity, the dispersion relation for a free tachyon becomes \ $E^{2}-{%
\mbox{\boldmath $p$}}^{2}=-m_{{\rm o}}^{2}$, and there is {\em no} need
therefore of imaginary masses...].

\ 

{\bf B)} \ {\bf Galactic Micro-quasars} -- Second: As to the {\em apparent}
Superluminal expansions observed in the core of quasars[8] and, recently, in
the so-called galactic microquasars[9], we shall not deal here with that
problem, too far from the other topics of this paper: without mentioning
that for those astronomical observations {\em there exist orthodox
interpretations, based on ref.}[10], {\em that are accepted by the
astrophysicists' majority.} \ For a theoretical discussion, see ref.[11].
Here, let us mention only that simple geometrical considerations in
Minkowski space show that a {\em single} Superluminal light source would
look[11,3]: \ (i) initially, in the ``optical boom'' phase (analogous to the
acoustic ``boom'' produced by a plane travelling with constant supersonic
speed), as an intense source which appears suddenly; and that \ (ii)
afterwards seem to split into TWO objects receding one from the other with
speed \ $V>2c$.

\ 

{\bf C)} \ {\bf Evanescent waves and ``tunnelling photons''} -- Third:
Within quantum mechanics (and precisely in the {\em tunnelling} processes),
it had been shown that the tunnelling time ---firstly evaluated as a simple
``phase time'' and later on calculated through the analysis of the
wavepacket behaviour--- does not depend on the barrier width in the case of
opaque barriers (``Hartman effect'')[12]: which implies Superluminal and
arbitrarily large (group) velocities $V$ inside long enough barriers: see
Fig.3. \ Experiments that may verify this prediction by, say, electrons are
difficult. Luckily enough, however, the Schroedinger equation in the
presence of a potential barrier is mathematically identical to the Helmholtz
equation for an electromagnetic wave propagating e.g. down a metallic
waveguide along the $x$-axis: and a barrier height $U$ bigger than the
electron energy $E$ corresponds (for a given wave frequency) to a waveguide
transverse size lower than a cut-off value. A segment of undersized guide
does therefore behave as a barrier for the wave (photonic barrier)[13]: So
that the wave assumes therein ---like an electron inside a quantum
barrier--- an imaginary momentum or wave-number and gets, as a consequence,
exponentially damped along $x$. In other words, it becomes an {\em evanescent%
} wave (going back to normal propagation, even if with reduced amplitude,
when the narrowing ends and the guide returns to its initial transverse
size). \ Thus, a tunnelling experiment can be simulated[13] by having
recourse to evanescent waves (for which the concept of group velocity can be
properly extended[14]). And the fact that evanescent waves travel with
Superluminal speeds has been actually {\em verified} in a series of famous
experiments (cf. Fig.4).

\hspace*{5 ex} Namely, various experiments ---performed since 1992 onwards
by G.Nimtz at Cologne[15], by R.Chiao's and A.Steinberg's group at
Berkeley[16], by A.Ranfagni and colleagues at Florence[17], and by others at
Vienna, Orsay, Rennes[17]--- verified that ``tunnelling photons" travel with
Superluminal group velocities. Such experiments roused a great deal of
interest[18], also within the non-specialized press, and were reported by
Scientific American, Nature, New Scientist, and even Newsweek, etc. \ Let us
add that also Extended Relativity had predicted[19] evanescent waves to be
endowed with faster-than-$c$ speeds; the whole matter appears to be
therefore theoretically selfconsistent. \ The debate in the current
literature does not refer to the experimental results (which can be
correctly reproduced by numerical elaborations[20,21] based on Maxwell
equations only), but rather to the question whether they allow, or do not
allow, sending signals or information with Superluminal speed[21,14].

\hspace*{5 ex} Let emphasize that the {\em most interesting} experiment of
this series is the one with two ``barriers" (e.g., with two segments of
udersized waveguide separated by a piece of normal-sized waveguide: Fig.5).
For suitable frequency bands ---i.e., for ``tunnelling" far from
resonances---, it was found that the total crossing time does not depend on
the length of the intermediate (normal) guide: namely, that the beam speed
along it is infinite[22]. \ This agrees with what predicted by Quantum
Mechanics for the non-resonant tunnelling trough two successive opaque
barriers (the tunnelling phase time, which depends on the entering energy,
has been shown by us to be {\em independent} of the distance between the two
barriers[23]). \ Such an important experiment could and should be repeated,
taking advantage also of the circumstance that quite interesting evanescence
regions can be easily constructed in the most varied manners, like by
several ``photonic band-gap materials" or gratings (it being possible tu use
from multilayer dielectric mirrors, to semiconductors, to photonic
crystals...)

\hspace*{5ex} We cannot skip a further topic ---which, being delicate,
should not appear in a brief review like this one--- since the last
experimental contribution to it (performed at Princeton by J.Wang et al. and
published in Nature on 7.20.00) is one of the two articles mentioned by the
N.Y.Times and commented at the end of July, 2000, by the whole world press.
\ Even if in Extended Relativity all the ordinary causal paradoxes seem to
be solvable[3,6], nevertheless one has to bear in mind that (whenever it is
met an object, ${\cal O}$, travelling with Superluminal speed) one may have
to deal with negative contributions to the tunnelling times[24]: and this
ought not to be regarded as unphysical. In fact, whenever an ``object''
(particle, electromagnetic pulse,,...) ${\cal O}$ {\em overcomes} the
infinite speed[3,6] with respect to a certain observer, it will afterwards
appear to the same observer as the ``{\em anti}-object'' $\overline{{\cal O}}
$ travelling in the opposite {\em space} direction[3,6]. \ For instance,
when going on from the lab to a frame ${\cal F}$ moving in the {\em same}
direction as the particles or waves entering the barrier region, the object $%
{\cal O}$ penetrating through the final part of the barrier (with almost
infinite speed[12,21,23], like in Figs.3) will appear in the frame ${\cal F}$
as an anti-object $\overline{{\cal O}}$ crossing that portion of the barrier 
{\em in the opposite space--direction\/}[3,6]. In the new frame ${\cal F}$,
therefore, such anti-object $\overline{{\cal O}}$ would yield a {\em negative%
} contribution to the tunnelling time: which could even result, in total, to
be negative. \ For any clarifications, see refs.[18]. \ Ci\`{o} che vogliamo
qui What we want to stress here is that the appearance of such negative
times is predicted by Relativity itself, on the basis of the ordinary
postulates[3,6,24,12,21]. \ (In the case of a non-polarized beam,, the wave
anti-packet coincides with the initial wave packet; if a photon is however
endowed with helicity $\lambda =+1$, the anti-photon will bear the opposite
helicity $\lambda =-1$). \ From the theoretical point of view, besides
refs.[24,12,21,6,3], see refs.[25]. \ On the (quite interesting!)
experimental side, see papers [26], the last one having already been
mentioned above.

\hspace*{5ex} Let us {\em add} here that, via quantum interference effects
in three-levels atomic systems, it is possible to obtain dielectrics with
refraction indices very rapidly varying as a function of frequency, with
almost complete absence of light absorption (i.e., with quantum induced
transparency) [27]. \ The group velocity of a light pulse propagating in
such a medium can decrease to very low values, either positive or negatives,
with {\em no} pulse distortion. \ It is known that experiments were
performed both in atomic samples at room temperature, and in Bose-Einstein
condensates, which showed the possibility of reducing the speed of light to
few meters per second. \ Similar, but negative group velocities ---implying
a propagation with Superluminal speeds thousands of time higher than the
previously mentioned ones--- have been recently predicted, in the presence
of such an ``electromagnetically induced transparency'', for light moving in
a rubidium condensate[28], while the corresponding experiments are being
dome at the Florence European laboratory ``LENS''.

\hspace*{5 ex} Finally, let us emphasize that faster-than-$c$ propagation of
light pulses can be (and was, in same cases) observed also by taking
advantage of anomalous dispersion near an absorbing line, or nonlinear and
linear gain lines, or nondispersive dielectric media, or inverted two-level
media, as well as of some parametric processes in nonlinear optics (cf.
G.Kurizki et al.)

\ 

{\bf D)} \ {\bf Superluminal Localized Solutions (SLS) to the wave
equations. The ``X-shaped waves"} -- The fourth sector (to leave aside the
others) is not less important. It returned after fashion when some groups of
capable scholars in engineering (for sociological reasons, most physicists
had abandoned the field) rediscovered by a series of clever works that any
wave equation ---to fix the ideas, let us think of the electromagnetic
case--- admit also solutions so much sub-luminal as Super-luminal (besides
the ordinary plane waves endowed with speed $c/n$). \ Let us recall that,
starting with the pioneering work by H.Bateman, it had slowly become known
that all homogeneous wave equations (in a general sense: scalar,
electromagnetic, spinorial,...) admit wavelet-type solutions with
sub-luminal group velocities[29]. \ Subsequently, also Superluminal
solutions started to be written down, in refs.[30] and, independently, in
refs.[31] (in one case just by the mere application of a Superluminal
Lorentz ``transformation"[3,32]).

\hspace*{5ex} A quite important feature of some new solutions of these
(which attracted much of the attention of the engineering colleagues) is
that they propagate as localized, non-dispersive pulses: namely, according
to the Courant and Hilbert's[29] terminology, as ``undistorted progressive
waves''. It is easy to realize the practical importance, for instance, of a
radio transmission carried out by localized beams, independently of their
being sub- or Super-luminal. \ But non-dispersive wave packets can be of use
also in theoretical physics for a reasonable representation of elementary
particles[33].

\hspace*{5ex} Within Extended Relativity since 1980 it had been found[34]
that ---whilst the simplest subluminal object conceivable is a small sphere,
or a point as its limit--- the simplest Superluminal objects results by
contrast to be (see refs.[34], and Figs.6 and 7) an ``X-shaped'' wave, or a
double cone as its limit, which moreover travels without deforming ---i.e.,
rigidly--- in a homogeneous medium[3]. \ It is worth noticing that the most
interesting localized solutions happened to be just the Superluminal ones,
and with a shape of that kind. \ Even more, since from Maxwell equations
under simple hypotheses one goes on to the usual {\em scalar} wave equation
for each electric or magnetic field component, one can expect the same
solutions to exist also in the field of acoustic waves, and of seismic waves
(and perhaps of gravitational waves too). \ Actually, such beams (as
suitable superpositions of Bessel beams) were mathematically constructed for
the first time, by Lu et al.[35], {\em in acoustics\/}: and were then called
``X-waves'' or rather X-shaped waves.

\hspace*{5ex} It is more important for us that the X-shaped waves have been
in effect produced in experiments both with acoustic and with
electromagnetic waves; that is, X-beams were produced that, in their medium,
travel undistorted with a speed larger than sound, in the first case, and
than light, in the second case. \ In Acoustics, the first experiment was
performed by Lu et al. themselves[36] in 1992, at the Mayo Clinic (and their
papers received the first 1992 IEEE award). \ In the electromagnetic case,
certainly more ``intriguing'', Superluminal localized X-shaped solutions
were first mathematically constructed (cf., e.g., Fig.8) in refs.[37], and
later on experimentally produced by Saari et al.[38] in 1997 at Tartu by
visible light (see fig.2 in ref.[38]!), and recently by Mugnai, Ranfagni
and Ruggeri at
Florence by microwaves[39] (this being the paper appeared in the Phys. Rev.
Lett. of May 22, 2000, which the national and international press refereed
to). \ Further experimental activity is in progress, for instance, at
Pirelli Cables, in Milan (by adopting as a source a pulsed laser) and at the
FEEC of Unicamp, Campinas, S.P.; while in the theoretical sector the
activity is even more intense, in order to build up ---for example--- new
analogous solutions with finite total energy or more suitable for high
frequencies, on one hand, and localized solutions Superluminally propagating
even along a normal waveguide[40], on the other hand.

\hspace*{5ex} Let us eventually touch the problem of producing an X-shaped
Superluminal wave like the one in Fig.7, but truncated ---of course-- in
space and in time (by the use of a finite, dynamic antenna, radiating for a
finite time): in such a situation, the wave will keep its localization and
Superluminality only along a certain ``depth of field'', decaying abruptly
afterwards[35,37]. \ We can become convinced about the possibility of
realizing it, by imaging the simple ideal case of a negligibly sized
Superluminal source $S$ endowed with speed $V>c$ in vacuum and emitting
electromagnetic waves $W$ (each one travelling with the invariant speed $c$%
). The electromagnetic waves will result to be internally tangent to an
enveloping cone $C$ having $S$ as its vertex, and as its axis the
propagation line $x$ of the source[3]. \ This is analogous to what happens
for a plane that moves in the air with constant supersonic speed. \ The
waves $W$ interfere negatively inside the cone $C$, and constructively only
on its surface. \ We can place a plane detector orthogonally to $x$, and
record magnitude and direction of the $W$ waves that hit on it, as
(cylindrically symmetric) functions of position and of time. \ It will be
enough, then, to replace the plane detector with a plane antenna which {\em %
emits} ---instead of recording--- exactly the same (axially symmetric)
space-time pattern of waves $W$, for constructing a cone-shaped
electromagnetic wave $C$ that will propagate with the Superluminal speed $V$
(of course, without a source any longer at its vertex): \ even if each wave $%
W$ travels with the invariant speed $c$. \ For further details, see the
first one of refs.[37]. \ Here let us only add that such localized
Superluminal waves appear to keep their good properties only as long as they
are fed by the waves arriving (with speed $c$) from the dynamic antenna:
taking the time needed for their generation into account, these waves seem
therefore unable to transmit information faster than $c$; however, they have
nothing to do with the illusory ``scissors effect'', since they certainly
carry energy-momentum Superluminally along their field depth (for instance,
they can get two detectors at a distance $L$ to click after a time {\em %
smaller} than $L/c$).

\hspace*{5ex} As we mentioned above, the existence of all these X-shaped
Superluminal (or ``Super-sonic'') seem to constitute at the
moment---together, e.g., with the Superluminality of evanescent waves--- one
of the best confirmation of Extended Relativity. \ It is curious than one of
the first applications of such X-waves (that takes advantage of their
propagation without deformation) is in progress in the field of medicine,
and precisely of ultrasound scanners[41]. \ A few years ago only, the
hypothesis that ``tachyons'' could be used to obtain directly 3-dimensional
ultrasound scans would have arisen the scepticism of any physicist, this
author included.\hfill \break 

\ 

{\bf Acknowledgments}

The author is deeply indebted to all the Organizers of this Conference, and
particularly to Larry Horwitz, J.D.Bekenstein and J.R.Fanchi, for their kind
invitation and warm, generous hospitality. For stimulating and friendly
discussions he is grateful to all the participants, and in particular to
J.D.Bekenstein, N.Ben-Amots, J.R.Fanchi, L.Horwitz, R.Lieu, M.Pav\v{s}i\v{c}%
. For further discussions or kind collaboration thanks are due also to
F.Bassani, A.Bertin, R.Chiao, A.Degli Antoni, F.Fontana, A.Gigli, H.E.Hern%
\'{a}ndez, G.Kurizki, J.-y.Lu, A.van der Merwe, D.Mugnai, G.Nimtz,
V.S.Olkhovsky, M.R.Zamboni, A.Ranfagni, R.A.Ricci, A.Shaarawi, D.Stauffer,
A.Steinberg, C.Vasini, M.T.Vasconselos and A.Vitale.

\ 

{\bf Figure captions}

\ 

Fig.1 -- Andamento dell'energia di un oggetto libero al variare della sua
velocit\`a.[2-4]\hfill\break

Fig.2 -- Illustrazione della ``regola di switching" (o principio di
reinterpretazione) di Stueckelberg-Feynman-Sudarshan[3-5]: $Q$ apparir\`a
essere l'antiparticella di P. \ Vedere il testo.\hfill\break

Fig.3 -- Andamento del ``tempo di penetrazione" di un pacchetto d'onde al
variare dello spazio percorso all'interno di una barriera di potenziale (da
Olkhovski, Recami, Raciti \& Zaichenko, ref.[12]). Secondo le predizioni
della meccanica quantistica, la velocit\`a all'interno della barriera cresce
illimitatamente per barriere opache; e il tempo di tunnelling non dipende
dalla larghezza della barriera[12].\hfill\break

Fig.4 -- Simulazione di tunnelling mediante esperimenti con onde evanescenti
(vedere il testo), le quali pure era previsto fossero Superluminali in base
alla Relativit\`a Estesa[3,4]. La figura mostra uno dei risultati delle
misure in refs.[15], ovvero la velocit\`a media di attraversamento della
regione di evanescenza (tratto di guida sottodimensionata, o ``barriera") al
variare della sua lunghezza. Come previsto[19,12], la velocit\`a media
supera $c$ per ``barriere" abbastanza lunghe.\hfill\break

Fig.5 -- L'interessante esperimento in guida d'onda metallica con due
barriere (tratti di guida sottodimensionata), cio\`e con due regioni di
evanescenza[22]. Vedere il testo.\hfill\break

Fig.6 -- Un oggetto intrinsecamente sferico (o al limite puntiforme) appare
come un elissoide contratto nella direzione del moto quando \`e dotato nel
vuoto di velocit\`a $v<c$. Qualora fosse dotato di velocit\`a $V>c$ (anche
se la barriera della velocit\`a $c$ non pu\`o essere attraversata n\'e da
sinistra n\'e da destra) apparirebbe[34] non pi\'u come una particella, ma
come un'onda ``a forma di X" che si disloca rigidamente (ovvero, come la
regione compresa tra un doppio cono e un iperboloide di rotazione a due
falde, o al limite come un doppio cono, che viaggia nel vuoto ---o in un
mezzo omogeneo--- Superluminalmente e senza deformazione).\hfill\break

Fig.7 -- Intersezioni con piani ortogonali alla direzione del moto di una
``X-shaped wave"[34], secondo la Relativit\`a Estesa[2-4]. L'esame della
figura suggerisce come costruire una semplice antenna dinamica atta a
generare tali onde Superluminali localizzate (una tale antenna fu in
effetti, indipendentemente, adottata da Lu et al.[36] per la prima
produzione di questi beams non-dispersivi).\hfill\break

Fig.8 -- Previsione teorica di onde Superluminali localizzate ``a forma di
X" per il caso elettromagnetico (da Lu, Greenleaf \& Recami[37], e
Recami[37]).\hfill\break

\newpage

{\bf Bibliography:}\hfill\break

[1] See, e.g., O.M.Bilaniuk, V.K.Deshpande \& E.C.G.Sudarshan: Am. J. Phys.
30 (1962) 718.

[2] See E.Recami \& R.Mignani: Rivista N. Cim. 4 (1974) 209-290; 4 (1974)
E398, and refs. therein. \ Cf. also E.Recami (editor): {\em Tachyons,
Monopoles, and Related Topics} (North-Holland; Amsterdam, 1978).

[3] E.Recami: Rivista N. Cim. 9 (1986), issue no.6 (pp.1$\div$178), and
refs. therein.

[4] See, e.g., E.Recami: in {\em Annuario 73, Enciclopedia EST}, ed. by
E.Macorini (Mondadori; Milano, 1973), pp. 85-94; \ and \ Nuovo Saggiatore 2
(1986), issue no.3, pp. 20-29.

[5] E.Recami: in {\em I Concetti della Fisica}, ed. by F.Pollini \&
G.Tarozzi (Acc.Naz.Sc.Lett.Arti; Modena, 1993), p.125-138; \ E.Recami \&
W.A.Rodrigues: ``Antiparticles from Special Found. Physics 12 (1982)
709-718; 13 (1983) E533.

[6] E.Recami: Found. Physics 17 (1987) 239-296. \ See also Lett. Nuovo
Cimento 44 (1985) 587-593; \ and P.Caldirola \& E.Recami: in {\em Italian
Studies in the Philosophy of Science}, ed. by M.Dalla Chiara (Reidel;
Boston, 1980), pp.249-298.

[7] Cf. M.Baldo Ceolin: ``Review of neutrino physics", invited talk at the 
{\em XXIII Int. Symp. on Multiparticle Dynamics (Aspen, CO; Sept.1993)}; \
E.W.Otten: Nucl. Phys. News 5 (1995) 11. \ From the theoretical point of
view, see, e.g., E.Giannetto, G.D. Maccarrone, R.Mignani \& E.Recami: Phys.
Lett. B178 (1986) 115-120 and refs. therein; \ S.Giani: ``Experimental
evidence of superluminal velocities in astrophysics and proposed
experiments", CP458, in {\em Space Technology and Applications International
Forum 1999}, ed. by M.S.El-Genk (A.I.P.; Melville, 1999), pp.881-888.

[8] See, e.g., J.A.Zensus \& T.J.Pearson (editors): {\em Superluminal Radio
Sources} (Cambridge Univ.Press; Cambridge, UK, 1987).

[9] I.F.Mirabel \& L.F.Rodriguez. : ``A superluminal source in the Galaxy",
Nature 371 (1994) 46 [with an editorial comment, ``A galactic speed record",
by G.Gisler, at page 18 of the same issue]; \ S.J.Tingay et al.:
``Relativistic motion in a nearby bright X-ray source", Nature 374 (1995)
141.

[10] M.J.Rees: Nature 211 (1966) 46; \ A.Cavaliere, P.Morrison \& L.Sartori:
Science 173 (1971) 525.

[11] E.Recami, A.Castellino, G.D.Maccarrone \& M.Rodon\`o: ``Considerations
about the apparent Superluminal expansions observed in astrophysics", Nuovo
Cimento B93 (1986) 119. \ Cf. also R.Mignani \& E.Recami: Gen. Relat. Grav.
5 (1974) 615.

[12] V.S.Olkhovsky \& E.Recami: Phys. Reports 214 (1992) 339, and refs.
therein, in particular T.E.Hartman: J. Appl. Phys. 33 (1962) 3427. \ See
also V.S.Olkhovsky, E.Recami, F.Raciti \& A.K.Zaichenko: J. de Phys.-I 5
(1995) 1351-1365.

[13] See, e.g., Th.Martin \& R.Landauer: Phys. Rev. A45 (1992) 2611; \
R.Y.Chiao, P.G.Kwiat \& A.M.Steinberg: Physica B175 (1991) 257; \
A.Ranfagni, D.Mugnai, P.Fabeni \& G.P.Pazzi: Appl. Phys. Lett. 58 (1991)
774; \ Y.Japha \& G.Kurizki: Phys. Rev. A53 (1996) 586. \ Cf. also
G.Kurizki, A.E.Kozhekin \& A.G.Kofman: Europhys. Lett. 42 (1998) 499: \
G.Kurizki, A.E.Kozhekin, A.G.Kofman \& M.Blaauboer: paper delivered at the
VII Seminar on Quantum Optics, Raubichi, BELARUS (May, 1998).

[14] E.Recami, F.Fontana \& R.Garavaglia: Int. J. Mod. Phys. A15 (2000)
2793, and refs. therein.

[15] G.Nimtz \& A.Enders: J. de Physique-I 2 (1992) 1693; \ 3 (1993) 1089; \
4 (1994) 1379; \ Phys. Rev. E48 (1993) 632; \ H.M.Brodowsky, W.Heitmann \&
G.Nimtz: J. de Physique-I 4 (1994) 565; \ Phys. Lett. A222 (1996) 125; A196
(1994) 154; \ G.Nimtz and W.Heitmann: Prog. Quant. Electr. 21 (1997) 81.

[16] A.M.Steinberg, P.G.Kwiat \& R.Y.Chiao: Phys. Rev. Lett. 71 (1993) 708,
and refs. therein; \ Scient. Am. 269 (1993) issue no.2, p.38. \ Cf. also
Y.Japha \& G.Kurizki: Phys. Rev. A53 (1996) 586.

[17] A.Ranfagni, P.Fabeni, G.P.Pazzi \& D.Mugnai: Phys. Rev. E48 (1993)
1453; \ Ch.Spielmann, R.Szipocs, A.Stingl \& F.Krausz: Phys. Rev. Lett. 73
(1994) 2308, \ Ph.Balcou \& L.Dutriaux: Phys. Rev. Lett. 78 (1997) 851; \
V.Laude \& P.Tournois: J. Opt. Soc. Am. B16 (1999) 194.

[18] Scientific American (Aug. 1993);\ Nature (Oct.21, 1993); \ New
Scientist (Apr. 1995); \ Newsweek (19 June 1995).

[19] Ref.[3], p.158 and pp.116-117. \ Cf. also D.Mugnai, A.Ranfagni,
R.Ruggeri, A.Agresti \& E.Recami: Phys. Lett. A209 (1995) 227.

[20] H.M.Brodowsky, W.Heitmann \& G.Nimtz: Phys. Lett. A222 (1996) 125.

[21] A.P.L.Barbero, H.E.Hern\'andez F., \& E.Recami: ``On the propagation
speed of evanescent modes" [LANL Archives \# physics/9811001], Phys. Rev.
E62 (2000) 8628, and refs. therein. \ See also E.Recami, H.E.Hern\'andez F.,
\& A.P.L.Barbero: Ann. der Phys. 7 (1998) 764-773.

[22] G.Nimtz, A.Enders \& H.Spieker: in {\em Wave and Particle in Light and
Matter}, ed. by A.van der Merwe \& A.Garuccio (Plenum; New York, 1993); \ J.
de Physique-I 4 (1994) 565. \ See also A.Enders \& G.Nimtz: Phys. Rev. B47
(1993) 9605.

[23] V.S.Olkhovsky, E.Recami \& G.Salesi: ``Tunneling through two successive
barriers and the Hartman (Superluminal) effect" [Lanl Archives \#
quant-ph/0002022], Report INFN/FM--00/20 (Frascati, 2000), submitted for
pub.; \ S.Esposito: (in preparazione).

[24] V.S.Olkhovsky, E.Recami, F.Raciti \& A.K.Zaichenko: ref.[12], pag.1361.
\ See also refs.[3,6], and E.Recami, F.Fontana \& R.Garavaglia: ref.[14],
pag.2807.

[25] R.Y.Chiao, A.E.Kozhekin A.E., and G.Kurizki: Phys. Rev. Lett. 77 (1996)
1254; \ C.G.B.Garret \& D.E.McCumber: Phys. Rev. A1 (1970) 305.

[26] S.Chu \& Wong W.: Phys. Rev. Lett. 48 (1982) 738; \ M.W.Mitchell \&
R.Y.Chiao: Phys. Lett. A230 (1997) 133-138; \ G.Nimtz: Europ. Phys. J. B (to
appear as a Rapid Note); \ L.J.Wang, A.Kuzmich \& A.Dogariu: Nature 406
(2000) 277; \ further experiments are being developed, e.g., at Glasgow
[D.Jaroszynski, private commun.] with X rays.

[27] G.Alzetta, A.Gozzini, L.Moi \& G.Orriols: Nuovo Cimento B36B (1976) 5.

[28] M.Artoni, G.C.La Rocca, F.S. Cataliotti \& F. Bassani: Phys. Rev. A (in
press).

[29] H.Bateman: {\em Electrical and Optical Wave Motion} (Cambridge
Univ.Press; Cambridge, 1915); \ R.Courant \& D.Hilbert: {\em Methods of
Mathematical Physics} (J.Wiley; New York, 1966), vol.2, p.760; \
J.N.Brittingham: J. Appl. Phys. 54 (1983) 1179; \ R.W.Ziolkowski: J. Math.
Phys. 26 (1985) 861; \ J.Durnin: J. Opt. Soc. 4 (1987) 651; \ A.O.Barut et
al.: Phys. Lett. A143 (1990) 349; \ Found. Phys. Lett. 3 (1990) 303; \
Found. Phys. 22 (1992) 1267.

[30] J.A.Stratton: {\em Electromagnetic Theory} (McGraw-Hill; New York,
1941), p.356; \ A.O.Barut et al.: Phys. Lett. A180 (1993) 5; A189 (1994) 277.

[31] R.Donnelly \& R.W.Ziolkowski: Proc. Roy. Soc. London A440 (1993) 541; \
I.M.Besieris, A.M.Shaarawi \& R.W.Ziolkowski: J. Math. Phys. 30 (1989) 1254;
\ S.Esposito: Phys. Lett. A225 (1997) 203; \ J.Vaz \& W.A.Rodrigues; Adv.
Appl. Cliff. Alg. S-7 (1997) 457.

[32] See also E.Recami \& W.A.Rodrigues Jr.: ``A model theory for tachyons
in two dimensions", in {\em Gravitational Radiation and Relativity}, ed. by
J.Weber \& T.M.Karade (World Scient.; Singapore, 1985), pp.151-203, and
refs. therein.

[33] A.M.Shaarawi, I.M.Besieris and R.W.Ziolkowski: J. Math. Phys. 31 (1990)
2511, Sect.VI; \ Nucl Phys. (Proc.Suppl.) B6 (1989) 255; \ Phys. Lett. A188
(1994) 218. \ See also V.K.Ignatovich: Found. Phys. 8 (1978) 565; and
A.O.Barut: Phys. Lett. A171 (1992) 1; A189 (1994) 277; Ann. Foundation L. de
Broglie, Jan.1994; and ``Quantum theory of single events: Localized de
Broglie--wavelets, Schroedinger waves and classical trajectories", preprint
IC/90/99 (ICTP; Trieste, 1990).

[34] A.O.Barut, G.D.Maccarrone \& E.Recami: Nuovo Cimento A71 (1982) 509; \
P.Caldirola, G.D.Maccarrone \& E.Recami: Lett. Nuovo Cim. 29 (1980) 241; \
E.Recami \& G.D.Maccarrone: Lett. Nuovo Cim. 28 (1980) 151.

[35] J.-y.Lu \& J.F.Greenleaf: IEEE Trans. Ultrason. Ferroelectr. Freq.
Control 39 (1992) 19.

[36] J.-y.Lu \& J.F.Greenleaf: IEEE Trans. Ultrason. Ferroelectr. Freq.
Control 39 (1992) 441.

[37] E.Recami: Physica A252 (1998) 586; \ J.-y.Lu, J.F.Greenleaf \&
E.Recami: ``Limited diffraction solutions to Maxwell (and Schroedinger)
equations'' [Lanl Archives \# physics/9610012], Report INFN/FM--96/01
(I.N.F.N.; Frascati, Oct.1996); \ R.W.Ziolkowski, I.M.Besieris \&
A.M.Shaarawi, J. Opt. Soc. Am. A10 (1993) 75.

[38] P.Saari \& K.Reivelt: ``Evidence of X-shaped propagation-invariant
localized light waves", Phys. Rev. Lett. 79 (1997) 4135-4138.

[39] D.Mugnai, A.Ranfagni and R.Ruggeri: {\em Phys. Rev. Lett.} 84 (2000)
4830.

[40] M.Z.Rached, E.Recami \& H.E.Hern\'{a}ndez-Figueroa: (in preparation); \
M.Z.Rached, E.Recami \& F.Fontana: ``Localized Superluminal solutions to
Maxwell equations propagating along a normal-sized waveguide'' [Lanl
Archives \# physics/0001039], subm.for pub.; \ I.M.Besieris, M.Abdel-Rahman,
A.Shaarawi \& A.Chatzipetros: Progress in Electromagnetic Research (PIER) 19
(1998) 1-48.

[41] J.-y.Lu, H.-h.Zou \& J.F.Greenleaf: Ultrasound in Medicine and Biology
20 (1994) 403; \ Ultrasonic Imaging 15 (1993) 134.

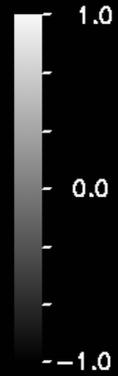
\end{document}